\documentclass[a4paper,12pt]{article}
\usepackage[round]{natbib}
\usepackage[english]{babel}
\usepackage{setspace}
\usepackage[utf8x]{inputenc}
\usepackage{graphicx} % Allows including images
\usepackage[left=2cm,right=2cm,top=2cm,bottom=2cm]{geometry}
\usepackage{textcomp}
\usepackage[T1]{fontenc}
\usepackage{amssymb}
\usepackage{xcolor}
\usepackage{authblk}
\usepackage{ulem}
\usepackage{hyperref}
\usepackage{lineno}
\usepackage{appendix}
\usepackage{float}
\usepackage{upgreek}
\title{
\Large
\textbf{The effect of near-surface winds on surface temperature\\ and dust transport on Venus}
}

\author[1]{Maxence Lef{\`e}vre}
\author[2]{S\'ebastien Lebonnois}
\author[2]{Aymeric Spiga}
\author[2]{Fran\c{c}ois Forget}
\affil[2]{LMD/IPSL, CNRS, Sorbonne Universit\'e, \'Ecole Normale Sup\'erieure, Universit\'e PSL, \'Ecole Polytechnique, Institut Polytechnique de Paris, Paris, France}
\affil[1]{LATMOS/IPSL, Sorbonne Universit\'e, UVSQ, Universit\'e Paris-Saclay, Centre National de la Recherche Scientifique, Paris, France}

\date{Published in JGR: Planets, 130, e2025JE009133. \url{https://doi.org/10.1029/2025JE009133}}

\begin{document}

\maketitle
%\newpage
\begin{abstract}
The knowledge of the Venus near-surface atmosphere is sparse. Few spacecrafts landed on the surface and measured winds with amplitudes below 1~m/s. The diurnal cycle of the wind amplitude and orientation is not known. Recent numerical simulations showed that slope winds along topographic structures could strongly impact the direction of winds. This study presents the first mesoscale modelling of such winds on Venus. A change of direction is occurring during the day in the main slopes, with upslope winds at noon due to solar heating and downslope winds at night. This is due to efficient IR cooling of the surface during the night, being colder than its surroundings slope atmospheric environment and leading to displacement of air. The temperature is impacted by the adiabatic cooling/warming induced by those winds. A strong heating effect is occurring for the downslope winds, leading to an anti-correlation between the surface temperature diurnal amplitude and the topography. This diurnal amplitude reaches 4~K in the plains and below 1~K in the mountains. The saltation of sediment by those winds was also quantified, with a higher probability at night along the slopes on the western flanks. 

\end{abstract}

\section*{Plain Language Summary}
Nighttime downslope winds are important meteorological features. It is the strongest land wind measured on Earth, whereas on Mars it is a key phenomenon of the dynamics of the near-surface region. On Venus, the importance of such is not known due to the lack of measurements at the surface. To investigate this topic, we use a regional model for the first time on Venus. During the day, there is a strong change of direction of these winds, going upslope at noon due to solar heating and downslope at night due to IR cooling. This change of direction particularly occurs near the Equator, where the solar flux is stronger. The temperature will also be impacted by adiabatic cooling/warming by those winds. A strong heating effect is associated with the downslope winds. The surface temperature diurnal amplitude is stronger in the plain than in the mountains, reaching respectively 4~K versus below 1~K. The surface temperature is therefore not only controlled by elevation but also by the slope. The lifting of sediment was also quantified with our model. A higher probability was estimated at night along the slopes. The western flank of the mountains would be preferable locations for sediment transport.

\section{Introduction}

The knowledge of the near-surface winds and temperature on Venus is essential, as the interaction between the surface and atmosphere is one of the objectives of future missions, with DAVINCI planning to measure temperature and winds near the surface \citep{garvinRevealingMysteriesVenus2022}, and VERITAS and EnVision observing aeolian features \citep{ghailVenSAREnVisionTaking2018}. In addition, several projects of long-duration landers at the surface of Venus are in development, like the Seismic and Atmospheric Exploration of Venus (SAEVe) concept \citep{kremicLongdurationVenusLander2020}. Characterization of the near-surface winds is also crucial to better understand the deep atmosphere dynamics and to constrain lander trajectories/landing \citep{knicelyStrategiesSafelyLanding2023}.

Only a limited number of probes have been able to collect data on the near-surface of Venus. At the surface, only VENERA-9 and 10 directly measured the wind for respectively 49 min and 90~s \citep{avduevskiiMeasurementWindVelocity1977}, and several other probes like VENERA-13 and 14 indirectly measured the wind speed \citep{ksanfomalitiWindVelocityVenus1983}. The amplitudes of the measured wind speeds are less than 2~m~s$^{-1}$ below 100~m \citep{lorenzSurfaceWindsVenus2016}, and typical values are below 0.5~m~s$^{-1}$ below 1~m. The diurnal cycle of the amplitude and orientation of the wind near the surface of Venus is not known, nor are the effects of the topography. The diurnal amplitude of the surface temperature can potentially impact the gas buffering at the surface and have implications on the surface evolution \citep{jakoskyBufferingDiurnalTemperature1984}. Due to the thick clouds and hot temperatures, remote surface temperature measurements are scarce. The surface temperature diurnal variation has been calculated using a variety of methods. It was estimated to be below 1~K based on radiative transfer modelling \citep{lewisVenusSurfaceTemperature1971,stoneDynamicsAtmosphereVenus1975}, around 15~K \citep{bohachevskyGeneralAtmosphericCirculation1973} and below 3~K \citep{lebonnoisPlanetaryBoundaryLayer2018} with GCM simulations, and about 6~K due to Planetary Boundary Layer dynamics \citep{gieraschGeneralCirculationVenus1997}. Two dune fields have been observed with Magellan radar measurements \citep{greeleyAeolianFeaturesVenus1992}, although the lack of knowledge about the spatial and temporal distribution of winds complicates the interpretation of sediment transport.

The Venus Planetary Climate Model (Venus PCM, formerly called the IPSL Venus GCM) simulations showed that the diurnal cycle of the planetary boundary layer (PBL) activity is correlated with the diurnal cycle of surface winds \citep{lebonnoisPlanetaryBoundaryLayer2018}. The global model predicted observed katabatic winds at night and anabatic winds during the day along the slopes of high-elevation terrains, resulting in a deeper PBL depth at midday.

Katabatic winds are an atmospheric flow formed when cooled dense air is accelerated downhill by gravity, overcoming the opposing along-slope pressure gradient \citep{mahrtRelationSlopeWinds1990}. These winds are particularly strong over ice-covered Greenland and the Antarctic, where the near-surface temperature inversions in winter reach 25~K, reaching typically 20~m~s$^{-1}$ with gusts exceeding 35~m~s$^{-1}$ \citep{nylenClimatologyKatabaticWinds2004}, around four times the equivalent outside polar regions. On Mars, the diurnal cycle of the surface temperature is three times larger than on Earth due to low thermal inertia and short radiative timescales of the thin CO$_2$ atmosphere, causing rapid transition from afternoon superadiabatic gradients to nighttime super-stable inversion and leading to winds up to 40~m~s$^{-1}$ \citep{richardsonPlanetWRFGeneralPurpose2007,spigaNewModelSimulate2009,richardsonRelationshipSurfacePressure2018} with typical temperature inversion between 20 and 30~K \citep{spigaElementsComparisonMartian2011}.

In this study, for the first time, we use mesoscale modelling to simulate the diurnal cycle of Venus' near-surface dynamics and especially estimate the influence of the topography on the surface temperature and saltation. Regional modelling has been extensively used to study these topics for the Earth \citep{schmidliAccuracySimulatedDiurnal2018,mikkolaDaytimeAlongvalleyWinds2023} and Mars \citep{rafkinMeteorologyGaleCrater2016,spigaKatabaticJumpsMartian2018,hernandez-bernalExtremelyElongatedCloud2022,langeModelingSlopeMicroclimates2023,montlaurThermallyDrivenWinds2024}.

In Section~\ref{Sec:Model1}, the model is described. In Section~\ref{Sec:Wind}, the impact of topography on the near-surface wind amplitude and direction is discussed. In Section~\ref{Sec:Temp}, the effect of the slope winds on the surface temperature is assessed. The contribution of these winds to the transport of sediment is estimated in Section~\ref{Sec:Dust}. Our conclusions are summarized in Section~\ref{Sec:Conc}.

\section{Modelling}
\label{Sec:Model1}

The LMD Venus mesoscale model \citep{lefevreMesoscaleModelingVenus2020} is based on the dynamical core of the Advanced Research Weather-Weather Research and Forecast (hereinafter referred to as WRF) terrestrial model \citep{skamarockTimesplitNonhydrostaticAtmospheric2008}. The WRF dynamical core integrates the fully compressible non-hydrostatic Navier-Stokes equations over a specified area of the planet and uses mass-coupled atmospheric variables (winds and potential temperature) with an explicitly conservative flux-form formulation of the fundamental equations to ensure the conservation of mass, momentum, and entropy \citep{skamarockTimesplitNonhydrostaticAtmospheric2008}

The solar and IR heating rates are calculated using the Venus PCM radiative transfer scheme \citep{lebonnoisAnalysisRadiativeBudget2015} and \citep{hausRadiativeHeatingCooling2015} solar rates look-up tables. The horizontal grid spacing for a mesoscale simulation is generally set to several tens of kilometers, and the convective turbulence in the planetary boundary layer and the cloud layer is not resolved. Therefore, similarly to what is done in global modelling, the LMD Venus mesoscale model uses a subgrid-scale parameterization for turbulent mixing. As in the Venus PCM, for mixing by smaller-scale turbulent eddies, the formalism of level 2.5 of \cite{mellorDevelopmentTurbulenceClosure1982} is adopted, which calculates with a prognostic equation the turbulent kinetic energy and mixing length. This scheme is taken from the LMDz model and is fully described in Appendix B of \cite{hourdinParameterizationDryConvective2002}. It has been used extensively in Martian conditions \citep{forgetImprovedGeneralCirculation1999}, computing temperature and wind with consistent in situ measurements. For Venus, no measurements exist to tune this scheme, but the temperature and winds close to the surface are qualitatively consistent with calculations in the PCM \citep{lebonnoisPlanetaryBoundaryLayer2018}. A simple dry convective adjustment is used to compute mixed layers in convectively unstable temperature profiles.

\subsection*{Simulation settings}

\begin{figure}[!ht]
  \centering
  \includegraphics[width=\textwidth]{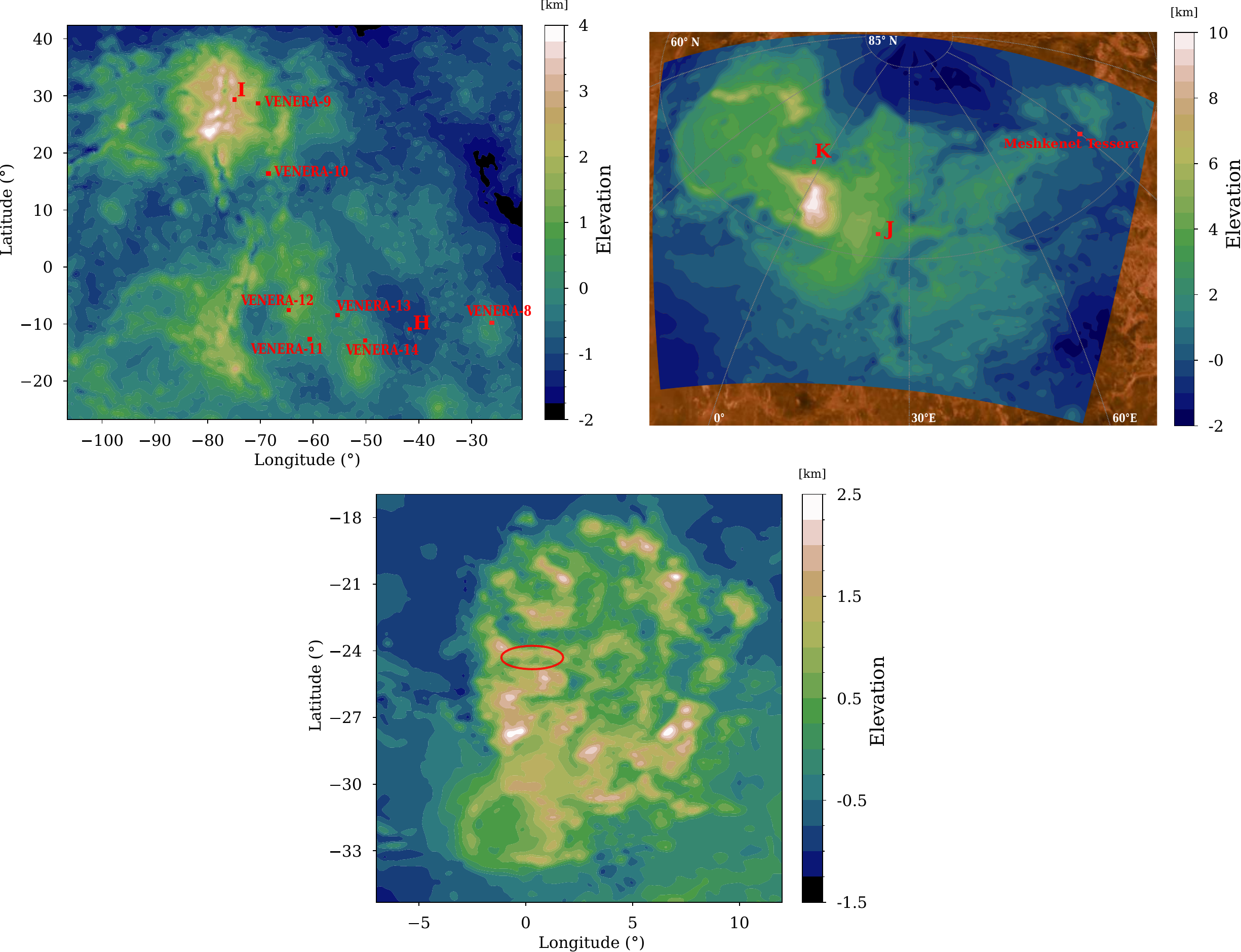}
    \caption{Elevation map of the Equatorial (top-left) and Polar (top-right) and Alpha Regio (bottom) domains with the location of the seven VENERA spacecrafts landing sites and 5 other points of interest represented by red squares. The red ellipse represents the landing area of DAVINCI \citep{garvinRevealingMysteriesVenus2022}.}
  \label{21}
\end{figure}

The three selected spatial domains shown in Fig~\ref{21} were built using Magellan topography data \citep{fordVenusTopographyKilometerscale1992} and Pioneer Venus data \citep{pettengillPioneerVenusRadar1980}. The Equatorial domain (top-left) was chosen to incorporate the seven VENERA landing sites considered in this study, all located in a rather confined location, within 80 000~km$^{2}$. The characteristics of these landing sites are summarized in Table~\ref{22}. The Polar domain (top-right) was chosen to incorporate Ishtar Terra, where Maxwell Montes, the highest mountain of Venus, is present with 10~km of altitude and slopes reaching 45$^{\circ}$. This domain was calculated using polar projection. The third domain (bottom) is centred over Alpha Regio, where DAVINCI will land \citep{garvinRevealingMysteriesVenus2022}. To have an accurate resolution of the slopes with a reasonable computational time, the horizontal resolution of the Equatorial domain is set to 20~km, 15~km for the Polar domain and 5~km for Alpha Regio. Given the horizontal resolutions involved, the dynamical timestep for mesoscale integrations is set to 8~s to ensure numerical stability. The vertical grid is the same as in \cite{lefevreMesoscaleModelingVenus2020}, with 150 levels from the surface to 90~km.

The settings of the simulations shown in this study are similar to the ones in \cite{lefevreMesoscaleModelingVenus2020}. The albedo, thermal inertia, and aerodynamic roughness height are set constant over the whole domain at respectively 0.5, 2000~J~m$^{-2}$~s$^{-1/2}$~K$^{-1}$ and 0.01~m. An 11-layer soil model is for the thermal conduction in the soil \citep{hourdinMeteorologicalVariabilityAnnual1993}. The horizontal boundary conditions have been chosen as 'specified', i.e. meteorological fields are extracted from the Venus PCM simulations \citep{garate-lopezLatitudinalVariationClouds2018} and interpolated both on the vertical grid, accounting for the refined topography of the mesoscale domain, and on the temporal dimension, accounting for the evolution of those fields over the low dynamical timestep in mesoscale integrations, ensuring the planetary-scale super-rotation of the Venusian atmosphere is well represented. The boundary fields are updated at a frequency of a 1/100 Venus day, enabling a correct representation of the large-scale variability simulated by Global Circulation Models (GCM) at the boundaries of the mesoscale domain. Between the mesoscale domain and the specified boundary fields, a relaxation zone set to 5 grid points is implemented to allow for the development of the mesoscale circulations inside the domain while keeping prescribed PCM fields at the boundaries \citep{skamarockTimesplitNonhydrostaticAtmospheric2008}. Simulations were performed for a whole Venus day, i.e. 117 Earth days.

\begin{table}[!ht]
\center
\scriptsize
\begin{tabular}{lcc}
\hline
Mission & Location & Measurements (m~s$^{-1}$) \\
\hline
VENERA-8 & 10.70$^{\circ}$S, 335.25$^{\circ}$E & 0.25~$\pm$~0.5$^a$ \\
& altitude : 193.8~m ; LT : early morning$^a$ & \\
\hline
VENERA-9 & 31.01$^{\circ}$N, 291.64$^{\circ}$E & 1.2~$\pm$~1.0$^a$ \\
& altitude : 1110~m ; LT : early afternoon$^a$ & 0.4–0.7$^b$ \\
\hline
VENERA-10 & 15.42$^{\circ}$N, 291.51$^{\circ}$E; & 0.6~$\pm$~1.0$^a$ \\
& altitude : -87.1~m ; LT : early afternoon$^a$ & 0.8–1.3$^b$ \\
\hline
VENERA-11 & 14$^{\circ}$S, 299$^{\circ}$E & $<$1.2$^c$ \\
& altitude : 234.1~m ; LT : mid-morning$^a$ & \\
\hline
VENERA-12 & 07$^{\circ}$S 294$^{\circ}$E ; & $<$1.2$^c$ \\
& altitude : 594.5~m ; LT : mid-morning$^a$ & \\
\hline
VENERA-13 & 07.05$^{\circ}$S 303$^{\circ}$E ; & 0.51(+0.19/-0.26) \\
& altitude : 429.6~m ; LT : mid-morning$^d$ & to 0.57(+0.43/-0.29)$^e$\\
\hline
VENERA-14 & 13.25$^{\circ}$S 310$^{\circ}$E ; & 0.35(+0.65/-0.18) \\
& altitude : 485.4 ; LT : mid-morning$^d$ & to 0.39(+0.61/-0.200)$^e$ \\
\hline
\end{tabular}
\smallbreak
\begin{minipage}{15cm}
\scriptsize
\item $^a$\citep{kerzhanovichAtmosphericDynamicsVenus1983} ; $^b$\citep{avduevskiiMeasurementWindVelocity1977} ; $^c$\citep{kerzhanovichVenera11Venera1980} \\ $^d$\citep{morozSummaryPreliminaryResults1983} ; $^e$\citep{ksanfomalitiWindVelocityVenus1983}
\end{minipage}
\medbreak
\caption{Summary of the VENERA 7 to 14 landing site informations and wind amplitude measurements.}
\label{22}
\end{table}
\normalsize

\section{Diurnal Variability of Wind Amplitude and Direction}
\label{Sec:Wind}

\begin{figure}[!ht]
  %\centeringmanuscript
  \includegraphics[width=16cm]{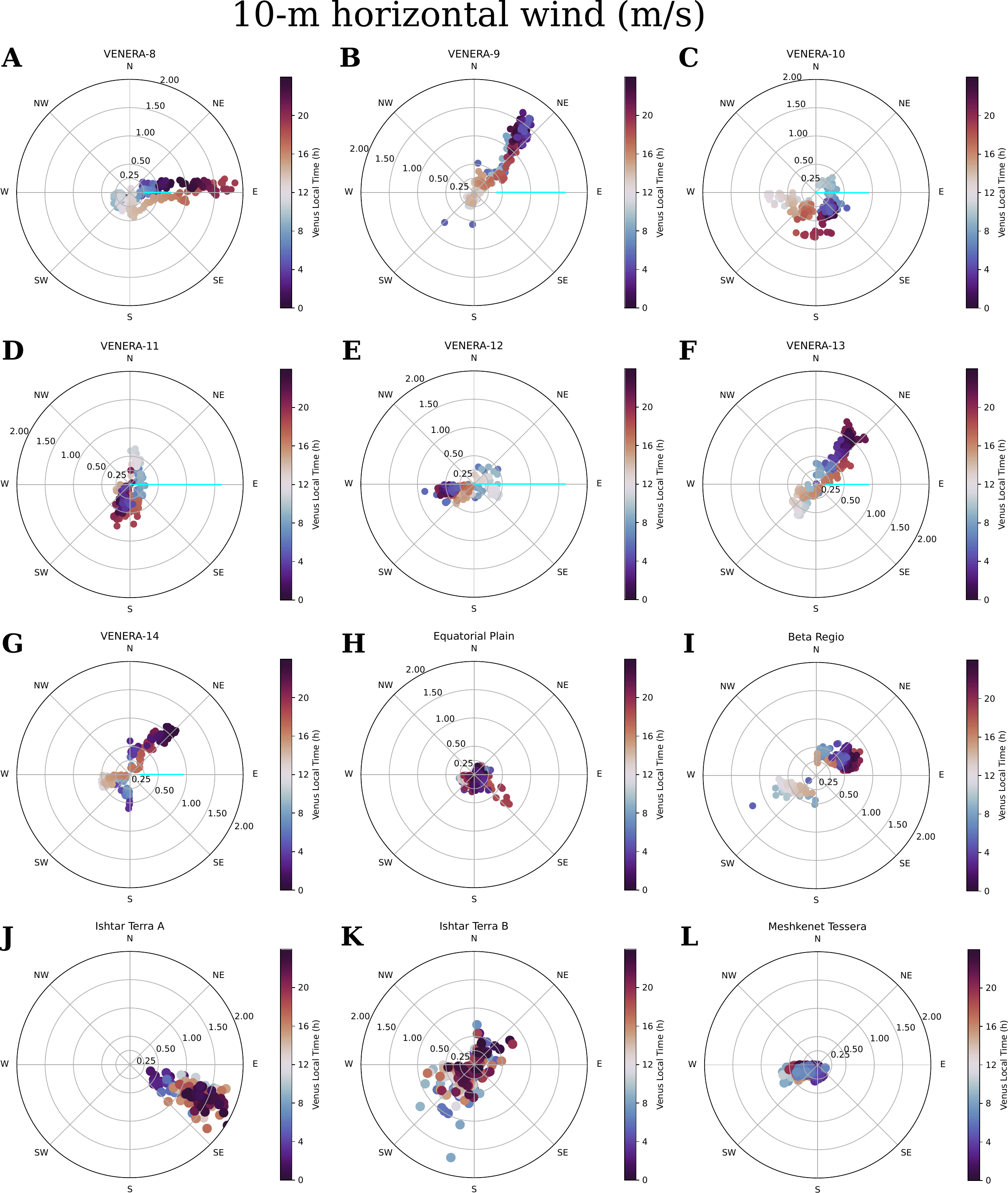}
    \caption{Sunflower wind chart of the 10~m horizontal wind for VENERA-8 (A), VENERA-9 (B), VENERA-10 (C), VENERA-11 (D), VENERA-12 (E), VENERA-13 (F) VENERA-14 (G) locations every 8 Venusian minutes (.i.e). H is a point in the Equatorial plain near Phoebe Regio, and I is a point in the East flank of Beta Regio. J and K are two points on Ishtar Terra. L is located in Meskhenet Tessera where a dune field was observed (See Fig~\ref{21}). The cyan lines represent the surface wind amplitude measured range for the VENERA landers in the late morning/early afternoon (Table~\ref {22}). No directions were estimated for VENERA observations, the Eastern direction is for illustration only.}
  \label{31}
\end{figure}

Fig~\ref{31} shows the sunflower wind chart of the 10~m horizontal wind (m~s$^{-1}$) for different locations pointed on Fig~\ref{21}. For the 7 VENERA landing sites, the horizontal wind amplitude resolved from the mesoscale model is below 1.6~m~s$^{-1}$ and around 0.5~m~s$^{-1}$ in the late morning/early afternoon window, consistent with the in-situ measurements all in this local time window (Table~\ref{22}). The amplitude of the wind varies during the day, except for the VENERA-10 location (C) and the Equatorial plain point (H). The landing site of VENERA-10 is between Phoebe and Beta Regio, a region where there is a strong convergence of winds. The wind is stronger at night and weaker around midday. For most of the points of interest, there is also a change in the wind direction, with a 180$^{\circ}$ shift between midnight and midday. The 7 VENERA landing points are located near mountains, where slope winds may impact the direction of winds. The VENERA-10, H and K locations have a more scattered change of direction during the day. Points I, J, and K are at higher altitudes than the other locations, resulting in stronger amplitudes. 

The amplitude distribution of the horizontal wind at 2, 10 and 100~m for the two domains at midday and midnight are shown in Fig~S1 and S2. The horizontal wind at 2~m (left) displays an amplitude between 0 and 1.5~m~s$^{-1}$, with 80$\%$ for both midday and midnight of its amplitude below 0.5~m~s$^{-1}$. At 10~m, winds can reach 1.75~m~s$^{-1}$, but with still 95$\%$ below 1~m~s$^{-1}$ at midday, against 90$\%$ at midnight. At 100~m, the wind's amplitude is increasing, reaching values as high as 3.0~m~s$^{-1}$. However, around 75$\%$ is below 1~m~s$^{-1}$ at midday and 60$\%$ at midnight. At midday, 95$\%$ of the horizontal wind is below 1.5~m~s$^{-1}$. These distributions at midday are consistent with the probability distribution estimated from the in-situ measurements at similar local time and latitude range \citep{lorenzSurfaceWindsVenus2016}. The difference between the two local times in the Equatorial region is representative of the diurnal cycle of the surface wind discussed in this study, with stronger wind at night. At 150~m of altitude, the large-scale winds reached 2~m~s$^{-1}$ in the Equatorial domain and 3~m~s$^{-1}$ in the North Pole, inferior by 25\% to the mesoscale winds. In Alpha Regio (Fig~S3), the distribution is similar to the Equatorial domain, with around 80$\%$ for both midday and midnight of its amplitude below 0.5~m~s$^{-1}$. However, the wind amplitude is slightly stronger by day due to the presence of  Lavinia Planitia in the vicinity of Alpha Regio, a 2000~km wide depression two kilometres below, where anabatic winds are heading toward the mountain.

\begin{figure}[!ht]
  \centering
  \includegraphics[width=12cm]{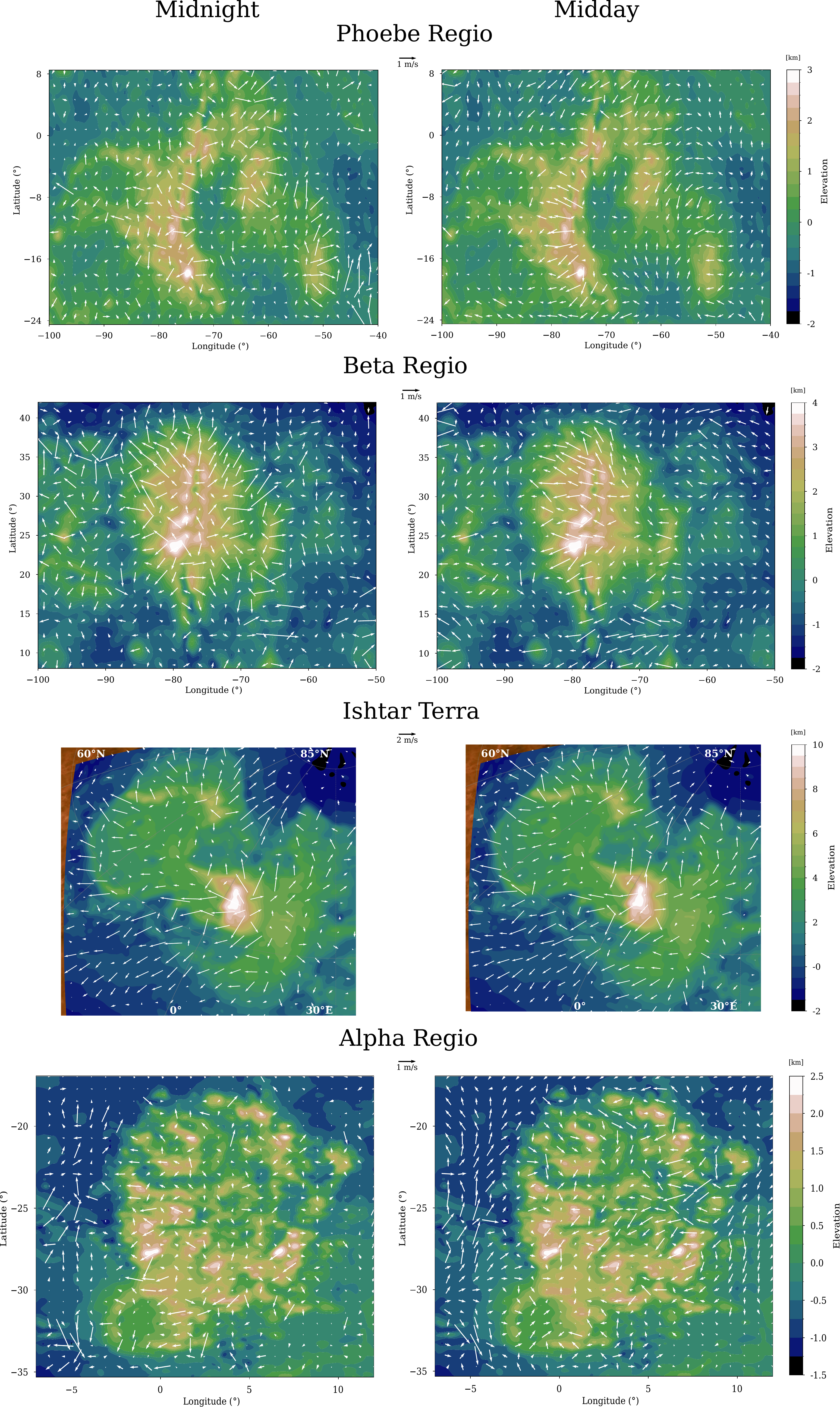}
    \caption{Snapshots maps of horizontal winds 10~m above the local surface (m~s$^{-1}$) at midnight (left column) and midday (right column) in the centre of the domain above Phoebe Regio, Beta Regio, Ishtar Terra and Alpha Regio.}
  \label{32}
\end{figure}

Fig~\ref{32} shows snapshots of the 10~m wind directions at midnight (left column) and midday (right column), focused on Phoebe Regio, Beta Regio, Ishtar Terra and the whole Alpha Regio domain. Maps of the entire Equatorial and Polar domain are shown in Fig~S4. In the first two lines in the Equator domain, the change of direction along a Venus day is visible with anabatic (upslope) winds at midday and katabatic (downslope) winds at midnight, especially on the eastern flanks of Beta Regio and Poebe Regio. This change of direction in the slope is due to a temperature inversion, specific to the environment. During the day, the solar flux is heating the flanks of a mountain, reaching a typical value at the Equator around 90~W~m$^{-2}$ at noon, consistent with in-situ measurements \citep{morozSummaryPreliminaryResults1983}. In this environment, this will lead to the surface of the mountain being hotter than its immediate atmospheric surroundings in the same isobar, and therefore less dense, leading to anabatic winds. At night, the surface is cooling more efficiently by the IR cooling than the atmosphere at the same isobar, typically -1$\cdot$10$^{-6}$~K~s$^{-1}$ on the slope against -1$\cdot$10$^{-7}$~K~s$^{-1}$ in the close atmosphere, leading to an immediate atmospheric environment colder and denser than its surroundings, engendering katabatic winds. This change of direction is consistent with the output of the Venus PCM \citep{lebonnoisPlanetaryBoundaryLayer2018}. Over Ishtar Terra, however, there are no changes of direction. The wind is constantly going downslope due to IR radiation cooling constantly during the day. The solar flux below 30~W~m$^{-2}$ on Ishtar Terra southern flank is too weak to engender anabatic winds. On Maxwell Montes, the highest mountain, the wind goes over the mountain throughout the day, reaching 3.0~m~s$^{-1}$. 

Several layers with significant decreases in surface emissivity and high radar reflectivities were observed at several altitudes on high-relief \citep{brossierLowRadarEmissivity2020}. The main hypothesis is the presence of strongly dielectric compounds, as ferroelectric materials \citep{treimanVenusRadarbrightHighlands2016} from atmosphere-surface interaction and/or precipitating minerals.
\cite{strezoskiSnowLineVenuss2022} reanalyzed Magellan data and showed that the emissivity drop altitudes vary over Maxwell Montes, suggesting that the atmosphere is progressively depleting in precipitable material as it crosses the mountain from south-east. 
The wind direction over Maxwell Montes is consistent with the atmospheric precipitate hypothesis, in addition to atmosphere–rock chemical reactions, to interpret Magellan radar features.

In Alpha Regio, there is also the presence of slope winds, katabatic by night and anabatic by day, especially on the boundary of the mountains. Due to the complex terrain, several points of wind convergence are present. The large-scale wind on the western side of the mountain is going east, contrary to the equatorial region, where the wind is globally western. This difference is due to Lavinia Planitia in the vicinity of Alpha Regio is driving the direction of the large-scale wind in this region. 

The amplitude of slope winds is typically around 1~m~s$^{-1}$ in the Equatorial region, one order of magnitude below its Earth and Martian equivalent, which reaches several tens of m~s$^{-1}$. From \cite{mcniderNoteVelocityFluctuations1982}, a frictionless upper limit value for the acceleration of the wind $\Gamma$ along a slope of inclination $\alpha$ is 

\begin{equation}
\Gamma = \frac{g sin\alpha \Delta T}{ 	\langle T \rangle}
\end{equation}
 \label{eq2}

\noindent where $\Delta T$ is the near-surface temperature inversion, $\langle T \rangle$ is the
average near-surface temperature and g is the acceleration of gravity equal to 8.87~m~s$^{-2}$. On Earth, typical radio-sounding profiles in the Sahara desert show at night a $\langle T \rangle$ equal to 285~K and a $\Delta T$ equal to 5~K between 0 and 250~m above the surface. For Mars at the same local time, $\langle T \rangle$ is equal to 175~K and $\Delta T$ 30~K between 0 and 250~m also at night \citep{spigaElementsComparisonMartian2011}. On Venus, on Beta Regio the typical temperature value is 705~K with a $\Delta T$ around 0.6~K. For the same inclination, the acceleration of the wind on Earth will be half that on Mars, but 20 times larger than on Venus. Therefore, due to its larger temperature and more stable atmosphere, the amplitude of slope winds on Venus is smaller. 

\section{Effect on Surface Temperature}
\label{Sec:Temp}

On Venus, the surface temperature is mainly determined by the elevation and is in thermal equilibrium with the atmosphere \citep{lecacheuxDetectionSurfaceVenus1993}, both due to the high density of the latter, the low solar flux reaching the surface, and the absence of latent heat flux. On Mars, the surface is, within good approximation, in radiative equilibrium with solar illumination because the atmosphere is so thin. However, the environment of slope winds departs from this equilibrium due to adiabatic heating/cooling and affects the surface temperature \citep{spigaImpactMartianMesoscale2011}.

\begin{figure}[!ht]
  \centering
  \includegraphics[width=16cm]{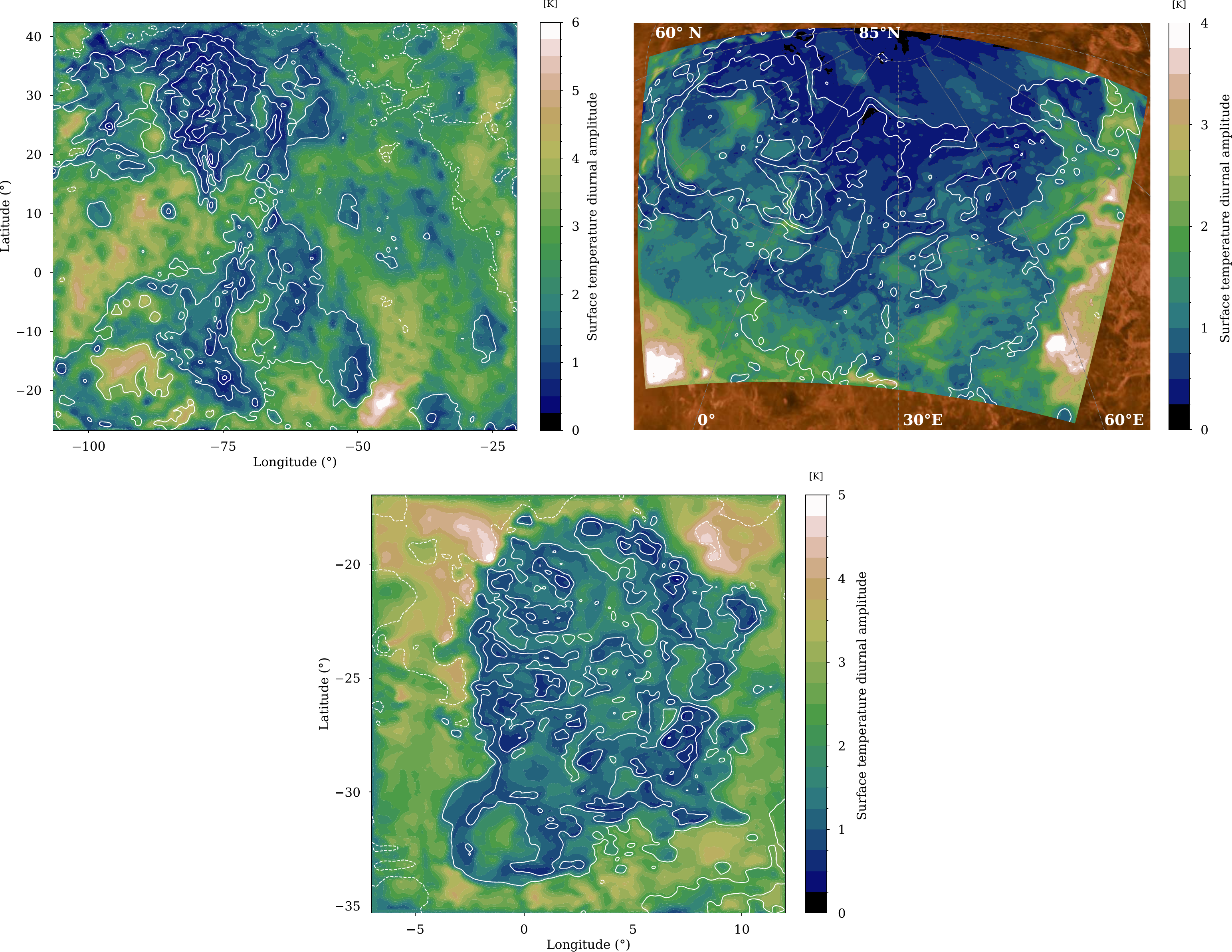}
    \caption{Map of the surface temperature diurnal amplitude (K) for Equatorial (top-left), Polar domain (top-right) and Alpha Regio (bottom). White contours represent the topography (Fig~\ref{21}), every kilometer for the Equatorial domain, two kilometers for the Polar domain and every 600~m for Alpha Regio.}
  \label{41}
\end{figure}

Fig~\ref{41} shows the surface temperature diurnal amplitude (K) for the three domains. The strong anti-correlation between surface temperature diurnal amplitude and topography is visible at the Equator and in Alpha Regio, where the low plains have a 4~K diurnal amplitude, whereas the high terrains have an amplitude below 1~K. In the Polar domain, there is no correlation between the topography and surface temperature diurnal amplitude. This amplitude is around 4~K in the low plains at 40$^\circ$ of latitude. At higher latitudes, the diurnal amplitude falls below 2~K due to the smaller solar heat flux.

\begin{figure}[!ht]
  \centering
  \includegraphics[width=16cm]{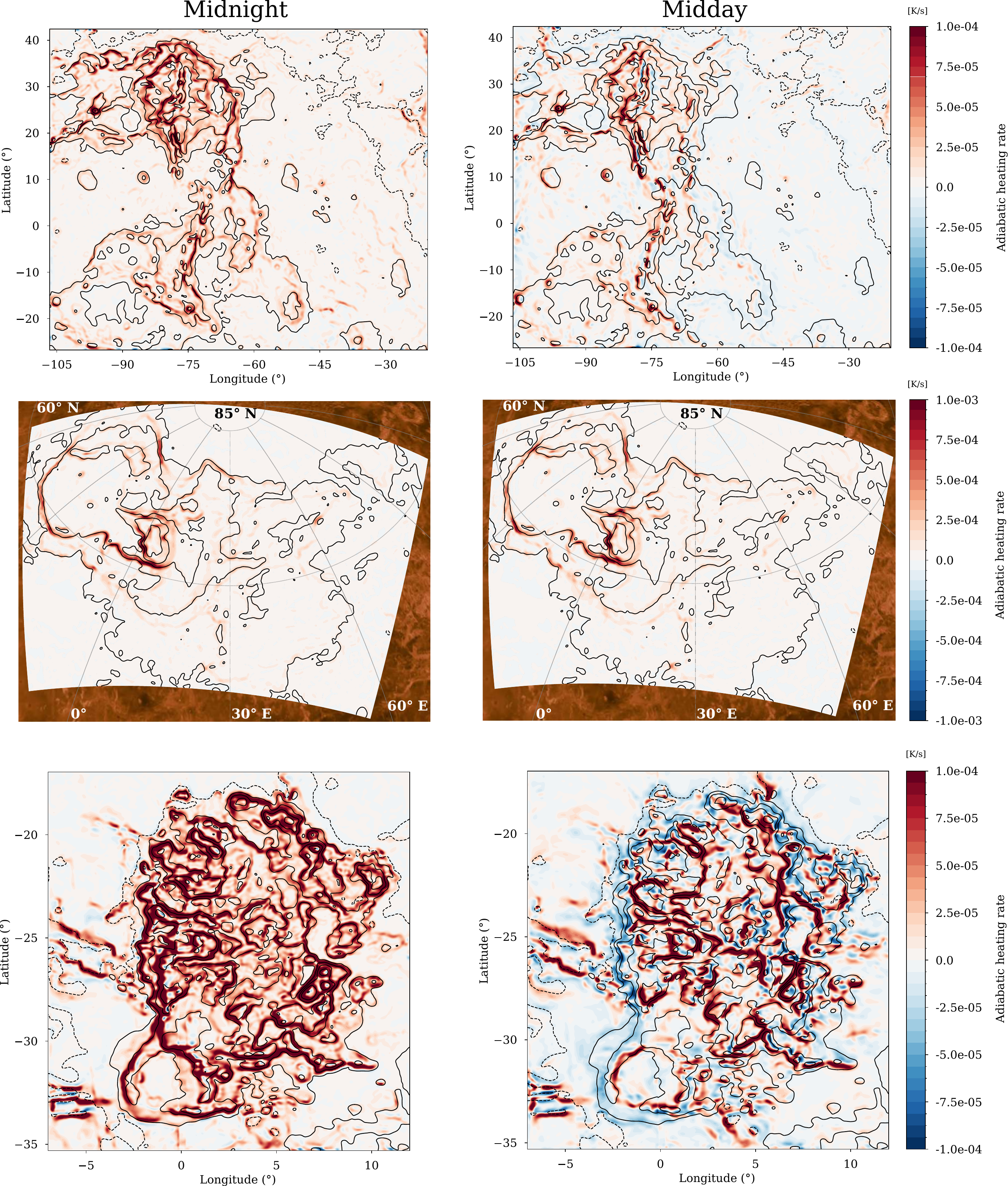}
    \caption{Top: Snapshots maps of the adiabatic heating rate of an atmospheric parcel with vertical wind velocity at 2~m above local surface $\jmath_{adiab}$ (K/s) at midnight (left) and midday (right) for the Equatorial (top) and Polar (centre) domains and Alpha Regio (bottom). Black contours represent the topography (Fig~\ref{21}), every kilometer for the Equatorial domain, two kilometers for the Polar domain and every 600~m for Alpha Regio.}
  \label{42}
\end{figure}

On the equatorial plains and outside Alpha Regio, the 3 to 4~K surface diurnal amplitude is due to the solar heating during the day and the nighttime IR cooling. In the mountainous region, there is a third contribution. To understand the impact of slope wind on surface temperature, the induced adiabatic cooling/warming is calculated as follows \citep{spigaImpactMartianMesoscale2011}:

\begin{equation}
\jmath_{adiab} = -\frac{g}{C_p}w
\end{equation}
 \label{eq1}

\noindent with g the acceleration of gravity, C$_p$ the heat capacity set to typical Venus tropospheric value of 1000~J~K$^{-1}$, and w the vertical wind component (m~s$^{-1}$) resolved by the model. Katabatic winds will heat the parcel, and anabatic winds will cool the parcel. Fig~\ref{42} shows the maps of $\jmath_{adiab}$ at 2~m for the three domains at midnight and midday.

The vertical component of the wind at 2~m has an average value of -0.002~m~s$^{-1}$ with a standard deviation of 0.002~m~s$^{-1}$, a maximum of 0.015~m~s$^{-1}$ and a minimum of -0.03~m~s$^{-1}$ for the Equatorial domain, averaged over the entire Venus day. For the Polar domain, the wind at 2~m has an average of -0.004~m~s$^{-1}$ with a standard deviation of 0.009~m~s$^{-1}$, a maximum of 0.02~m~s$^{-1}$ and a minimum of -0.2~m~s$^{-1}$. Diurnal variation of near-sruface vertical wind for Beta Regio and the equatorial plain, point I and H in Fig\ref{21}, is shown in Fig~S5.  With these values of vertical wind, the slope wind adiabatic heating rates reach values of -1$\cdot$10$^{-7}$ and 1$\cdot$10$^{-5}$~K~s$^{-1}$ in the Equatorial plain and Beta Regio respectively at night. Typical values of the nighttime longwave cooling rate on a slope are around -1$\cdot$10$^{-6}$~K~s$^{-1}$, i.e. an order of magnitude lower than the slope wind rates. The timescale of radiative cooling is therefore longer and cannot balance the warming induced by katabatic adiabatic compression through vertical motion, leading to a decrease of the surface temperature diurnal amplitude in the slopes compared to plains. 

This effect of slope on the temperature shows that near the Equator, the surface temperature is not only controlled by the elevation but also by the slope. This should be considered when measuring the surface properties like emissivity and thermal inertia. A Surface Brightness 4~K temperature anomaly was retrieved at the same latitude and altitude between a plain and a mountainous region \citep{muellerMultispectralSurfaceEmissivity2020}, possibly linked to a difference in surface temperature.

\section{Eolian transport}
\label{Sec:Dust}

Only two prominent dune fields have been identified on Venus with the Magellan radar \citep{greeleyAeolianFeaturesVenus1992}, Algaonice at 25$^{\circ}$S, 340$^{\circ}$E covers some 1300 km~$^2$ with bright dunes of 0.5-5~km in length with a wavelength around 0.5~km, and Fortuna-Meskhenet at 67$^{\circ}$N, 91$^{\circ}$E with transverse dunes of 0.5-10~km long, 0.2-0.5~km wide and spaced by an average of 0.5~km. Several sites with micro-dunes or small-scale wind streaks have also been proposed \citep{weitzDunesMicrodunesVenus1994,bondarenkoNorthsouthRoughnessAnisotropy2006}, but the resolution of the Magellan radar was not high enough to confirm them.

Within the lowest portion of the planetary boundary layer, a semi-empirical logarithmic wind profile approximated by the Prandlt-von K\'arm\'an equation defining the friction velocity u$^{\star}$ is commonly used to describe the vertical distribution of horizontal mean wind speed:
 \begin{equation}
u_{\star}~=~k \frac{u_z}{ln(z/z_0)}
\end{equation}
 \label{eq3}
With the Von K\'arm\'an constant, k, equal to 0.4, the aerodynamic roughness height, z$_0$, is set to 1~cm, in the range of values estimated from Magellan radar measurements \citep{blumbergVenusInfluenceSurface1994}. u$_z$ is the horizontal wind at the altitude z.

Theoretical calculations and laboratory experiments estimated the threshold friction velocity u$_{\star t}$ for which the dust is to be lifted in Venus surface conditions to be minimum around 2.5~10$^{-2}$~m~s$^{-1}$~m~s$^{-1}$ \citep{iversenWindblownDustEarth1976,iversenSaltationThresholdEarth1982,gunnConditionsAeolianTransport2022}. The stress associated $\tau$ = $\rho \cdot u^{2}_{\star t}$, with $\rho$ the density of the atmosphere, is about two times lower for Venus than for Earth, and three times lower than for Mars. Such a threshold value depends also on the density of the sediment, data that is not well-known yet. These thresholds are calculated for a 2.65~g~cm$^{-3}$ particle density, but no measurement exists for Venusian surface particles.

\begin{figure}[!ht]
  \centering
  \includegraphics[width=16cm]{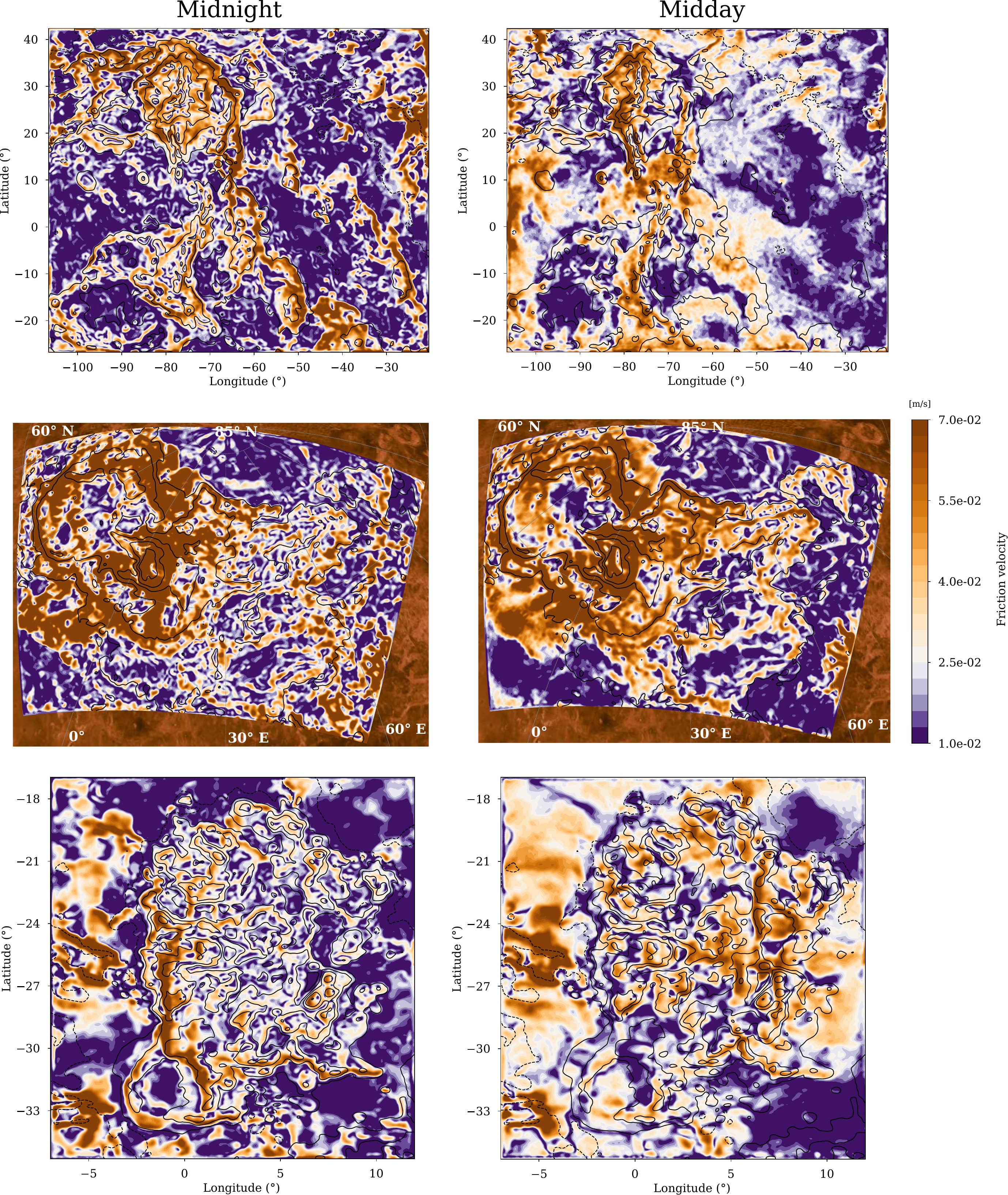}
    \caption{Map of the friction velocity u$^{\star}$ at midnight (left) and midday (right) for the Equatorial (top) and Polar (centre) domains and Alpha Regio (bottom). Back contours represent the topography (Fig~\ref{21}), every kilometer for the Equatorial domain, two kilometers for the Polar domain and every 600~m for Alpha Regio.}
  \label{51}
\end{figure}

Fig~\ref{51} shows instantaneous maps of u$^{\star}$ from the 2-m wind for the three domains at midnight and midday. For the Equatorial domains, the median value of u$^{\star}$ is around 2~10$^{-2}$~m~s$^{-1}$, several orders of magnitude below Earth and Mars \citep{gunnConditionsAeolianTransport2022}. The friction velocity is stronger at night due to higher horizontal wind amplitude. At midday, 33$\%$ of the domain have a u$^{\star}$ value above the 2.5~10$^{-2}$~m~s$^{-1}$ u$^{\star}_t$ for particle of 75~$\upmu$m, and 1$\%$ of the domain is above the 7~10$^{-2}$~m~s$^{-1}$ u$^{\star}_t$ for particle of 10~$\upmu$m or 1mm. At night, it reaches 36$\%$ of the domain above the 75~$\upmu$m u$^{\star}_t$ and 3$\%$ for 10~$\upmu$m (or 1mm) threshold. For the Polar domain, 43$\%$ at midday, and 46$\%$ at night of the domain above the 75~$\upmu$m threshold, against 18$\%$ at midday and 19$\%$ at night above the 10~$\upmu$m threshold. In the Equator, some locations are above the saltation threshold both at midnight and midday. This is the case for the western flank of the mountains due to the western direction of the large-scale wind. The region at 12$^{\circ}$N and -70$^{\circ}$ of longitude, between Phoebe Regio and Beta Regio, is also a region of interest for sedimentary transport, especially during the day, due to strong gap winds on the flat land between those two mountains (Fig~\ref{21}). In Alpha Regio, at midnight 29$\%$ of the domain have a u$^{\star}$ value above the 2.5~10$^{-2}$~m~s$^{-1}$ u$^{\star}_t$ for particle of 75~$\upmu$m, and 2$\%$ of the domain is above the 7~10$^{-2}$~m~s$^{-1}$ u$^{\star}_t$ for particle of 10~$\upmu$m or 1mm, whereas at midday 45$\%$ and 2$\%$ are respectively above the 2.5~10$^{-2}$~m~s$^{-1}$ and 7~10$^{-2}$~m~s$^{-1}$ threshold.

In Fortuna-Meskhenet region, where dune fields were observed, the friction velocity is at the 0.025~m~s$^{-1}$ threshold, meaning that dust can be lifted in the area by regional slope winds. The dune field is composed of two main directions, with a southeast-to-northwest wind flow shifting to a westward flow in the northern part. In Fig~\ref{31}-L, the wind is moving towards the west, consistent with the observed northern part. However, the resolution of the model is too coarse to resolve the complex terrain and account for the wind direction.

With the estimation of the friction velocity, information on the flow regime can be evaluated, and the wavelength of dunes can be estimated \citep{courrechdupontComplementaryClassificationsAeolian2024}. The Reynolds number $R_r$ = r$\cdot$u$_*$/$\upnu$ with r the aerodynamic roughness height fixed to 1~cm and $\upnu$ the viscosity of the atmosphere equal to 4$\cdot$10$^{-7}$~m$^2$~s$^{-1}$, is superior to 100 on 98$\%$ of the domain. The flow is then in an aerodynamically rough regime, meaning that the viscous sublayer is perturbed by the turbulence mixing caused by the aerodynamic roughness height. In this regime, the minimum wavelength of the dunes is found to scale with the saturation length $L_{sat}$, equal to 4.4$\cdot$d$\cdot$$\uprho_s$/$\uprho_f$ \citep{claudinScalingLawAeolian2006}, with d the radius of the particle and $\uprho_p$/$\uprho_f$ the ratio between the density of the particle and the density of the fluid. For the lowest saltation threshold, the particle radius is equal to 75$\upmu$m with a $\uprho_p$/$\uprho_f$ of 41 \citep{iversenSaltationThresholdEarth1982}, leading to a minimum wavelength of around 13~cm. The maximum wavelength of the dune would be of the order of the planetary boundary layer (PBL) height. No measurements exist of the turbulence at the surface of Venus. For the Venus PCM \citep{lebonnoisPlanetaryBoundaryLayer2018} and Large Eddy Simulations \citep{lefevreVenusBoundaryLayer2022}, on the Equatorial plains, the PBL height varies from a few hundred meters at night to around 1~km. On the high terrain, the PBL height can reach 7~km. At high latitudes, the PBL height would be below 1~km due to the weak solar flux. 

\section{Conclusion}
\label{Sec:Conc}
In the next decade, there will be two radars \citep{ghailVenSAREnVisionTaking2018} on board Envision and VERITAS that will monitor the surface composition and geomorphological features. This study provides for the first time at a mesoscale level the near-surface dynamics of the Venusian atmosphere. It is shown that the main elevation features in the tropics control the dynamics in this region and that their vicinities are preferable locations for the sediment transport on Venus. The amplitude of 2~m-wind is 80$\%$ below 0.5~m~s$^{-1}$ and 95$\%$ below 1~m~s$^{-1}$. This is consistent with the amplitude distribution deduced from measurements. 
The amplitude of the wind resolved by the model is in agreement with the VENERA measurements at the same local time. Due to the daytime solar heating and nighttime IR cooling, the wind amplitude and direction exhibit a clear diurnal cycle in the tropics, with the generation of slope winds. During the day, winds are going upslope, whereas at night they are going downslope.
At the Poles, the solar flux at the surface is too small to induce anabatic winds during the day, resulting in constant katabatic winds on the largest slopes. These slope winds engendered by the thermal balance of the slope surface are thought to be ubiquitous in the tropics due to the absence of moist processes and seasonal effects and the stability of the cloud coverage. \\

The present modeling study reports the first estimate for Venus of the impact of slope winds on surface temperature. The slope winds impact the surface temperature diurnal amplitude, anabatic winds will cool down the environment, while katabatic winds will heat it up. On the low Equatorial plains, the surface temperature will vary by 4~K. However, on the slopes, the katabatic winds will heat the surface at a rate that the IR cooling cannot balance, resulting in a diurnal amplitude below 1~K. On  Venus, the surface temperature in the tropics is therefore not only controlled by elevation but also by the slope. \\
The saltation of sediment by those winds was also estimated using the friction velocity formalism. The slope winds are dominating the near-surface dynamics, and a third of the domain in the tropics and almost half of the polar domain are above the lowest saltation threshold, i.e. for 75~$\upmu$m particles, meaning that mesoscale winds are strong enough to lift dust of that size. These regions are close to mountain flanks, but flatter regions where you have wind convergence are also regions of saltation. However, the sediment reservoir is not known. One of the two regions where dune fields were observed is at the lowest saltation threshold level. Based on the value of the friction velocity, the flow is in a rough regime, leading to dune wavelengths between 13~cm and the PBL height, strongly varying with latitude, local time and height.

Several developments can improve the knowledge of the impact of slope winds in the Venusian environment. The heat capacity is set constant by design in the model. However, CO$_2$ heat capacity strongly varies with temperature \citep{lebonnoisSuperrotationVenusAtmosphere2010}. Taking into account this peculiarity would provide a more realistic description of the stability of the atmosphere and the heating/cooling of slope winds.\\
The implementation of a thermal plume model to better represent subgrid turbulence, already in use for other planets, and tuned with future data, would be an additional refinement of the modelling.\\
This study considers surface properties such as emissivity, albedo, thermal inertia, and aerodynamic roughness height constant. Taking into account spatial variation of those characteristics depending on surface composition in future studies, like on Mars \citep{spigaNewModelSimulate2009}, would provide a more realistic description of the Venusian near-surface dynamics. \\
Simulations at higher resolution and with a dust tracer should be performed to be able to assess the dust transport realistically.

\section*{acknowledgments}
The authors would like to thank the two anonymous reviewers who helped to improve this study. This project has received funding from the European Union's Horizon Europe research and innovation program under the Marie Sk\l odowska-Curie grant agreement 101110489/MuSICA-V. ML would like to acknowledge the use of the Sorbonne Université SACADO service unit.

\section*{Open research}
The Simulation outputs used to obtain the results in this paper are available in the open online repository \citep{Lefe24data}.

\newpage
\appendix
\begin{appendices}
\section{Supporting Information}
\begin{figure}[!ht]
  \centering
  \includegraphics[width=16cm]{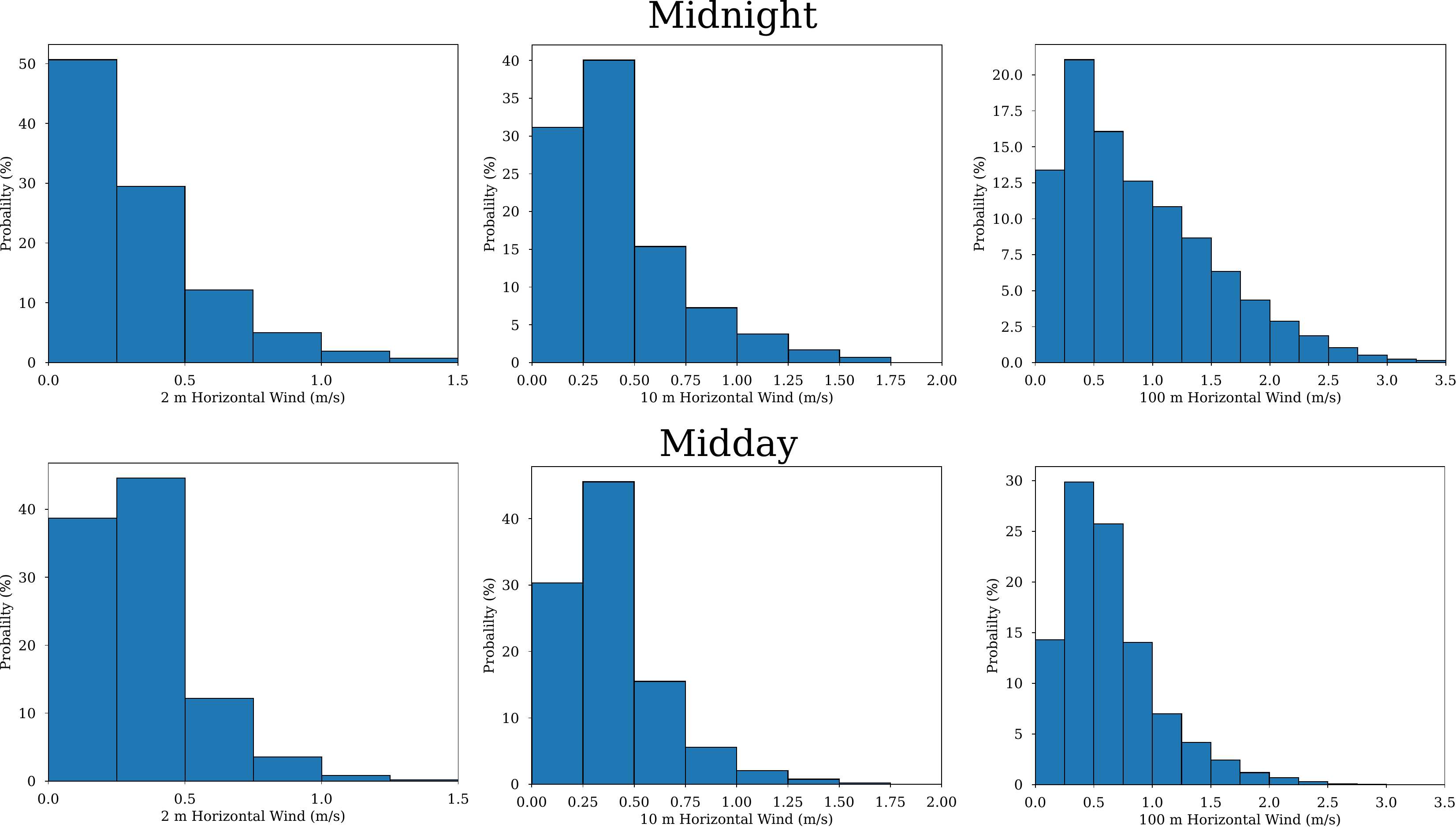}
    \caption{Histogram of horizontal wind amplitude distribution at 2~m (left), 10~m (centre) and 100~m (right) above local surface (m~s$^{-1}$) at midnight (top) and midday (bottom) for the Equatorial domain.}
  \label{A21}
\end{figure}

\begin{figure}[!ht]
  \centering
  \includegraphics[width=16cm]{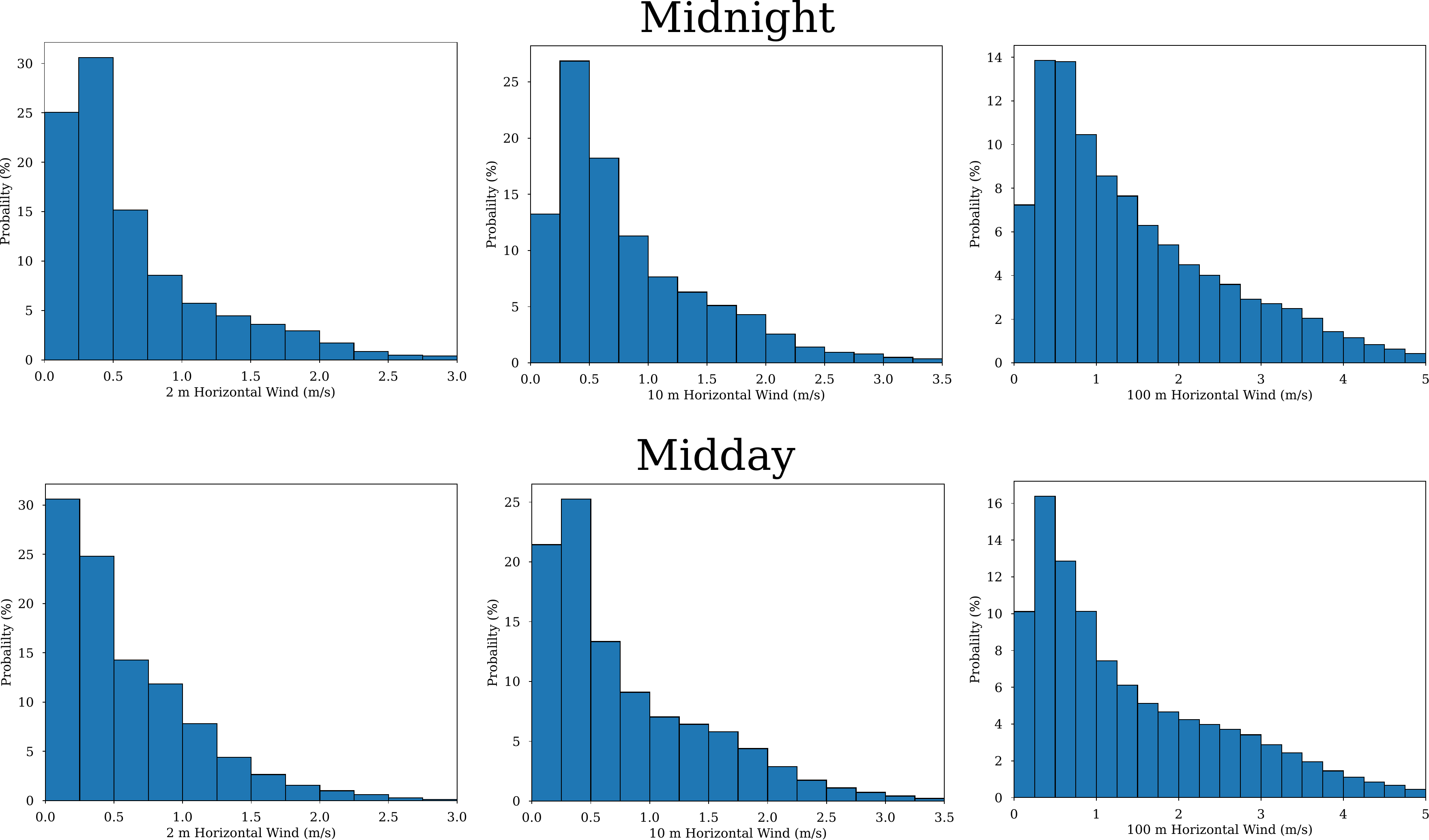}
    \caption{Histogram of horizontal wind amplitude distribution at 2~m (left), 10~m (centre) and 100~m (right) above local surface (m~s$^{-1}$) at midnight (top) and midday (bottom) for the Polar domain.}
  \label{A22}
\end{figure}

\begin{figure}[!ht]
  \centering
  \includegraphics[width=16cm]{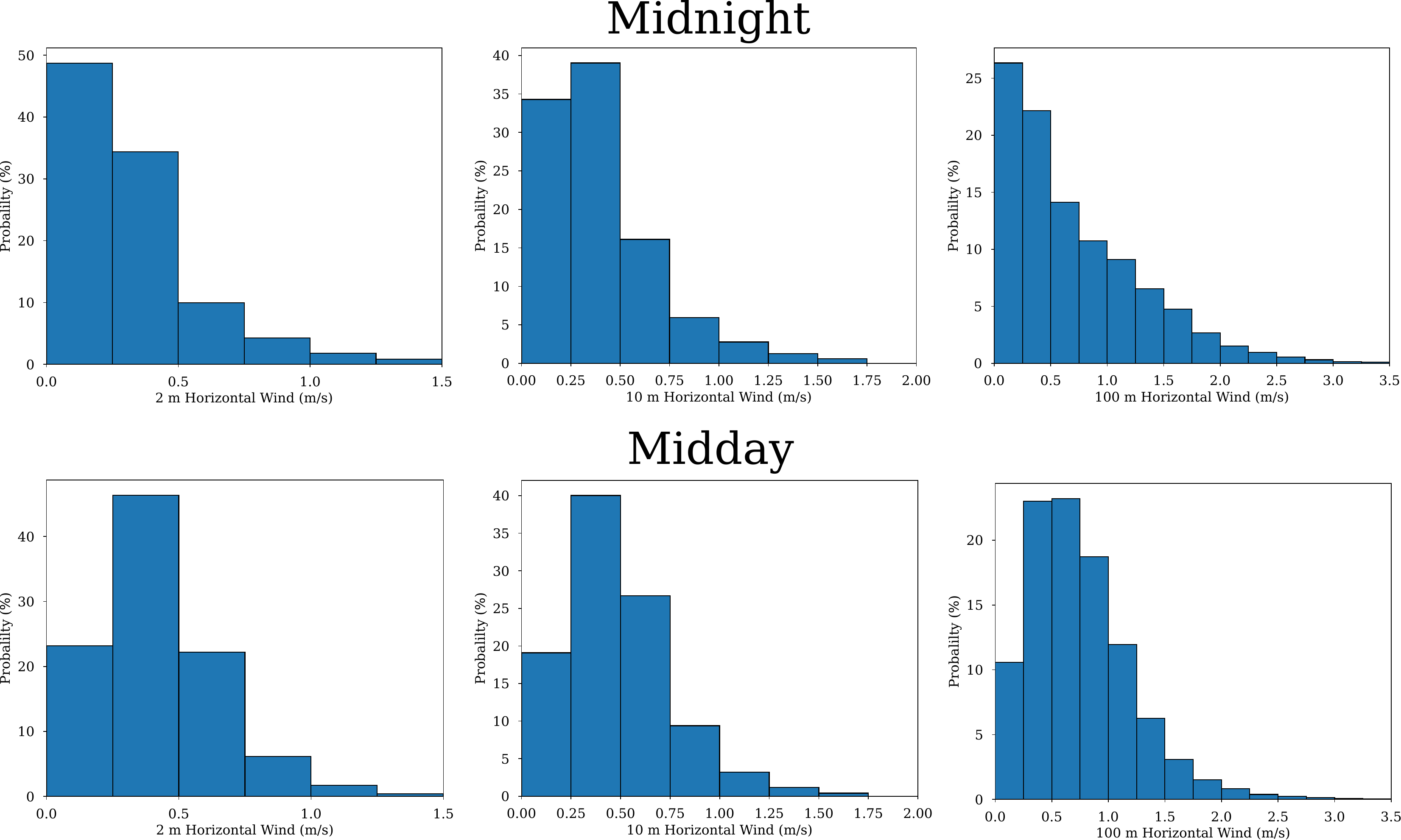}
    \caption{Histogram of horizontal wind amplitude distribution at 2~m (left), 10~m (centre) and 100~m (right) above local surface (m~s$^{-1}$) at midnight (top) and midday (bottom) for Alpha Regio.}
  \label{A22}
\end{figure}

\begin{figure}[!ht]
  \centering
  \includegraphics[width=15cm]{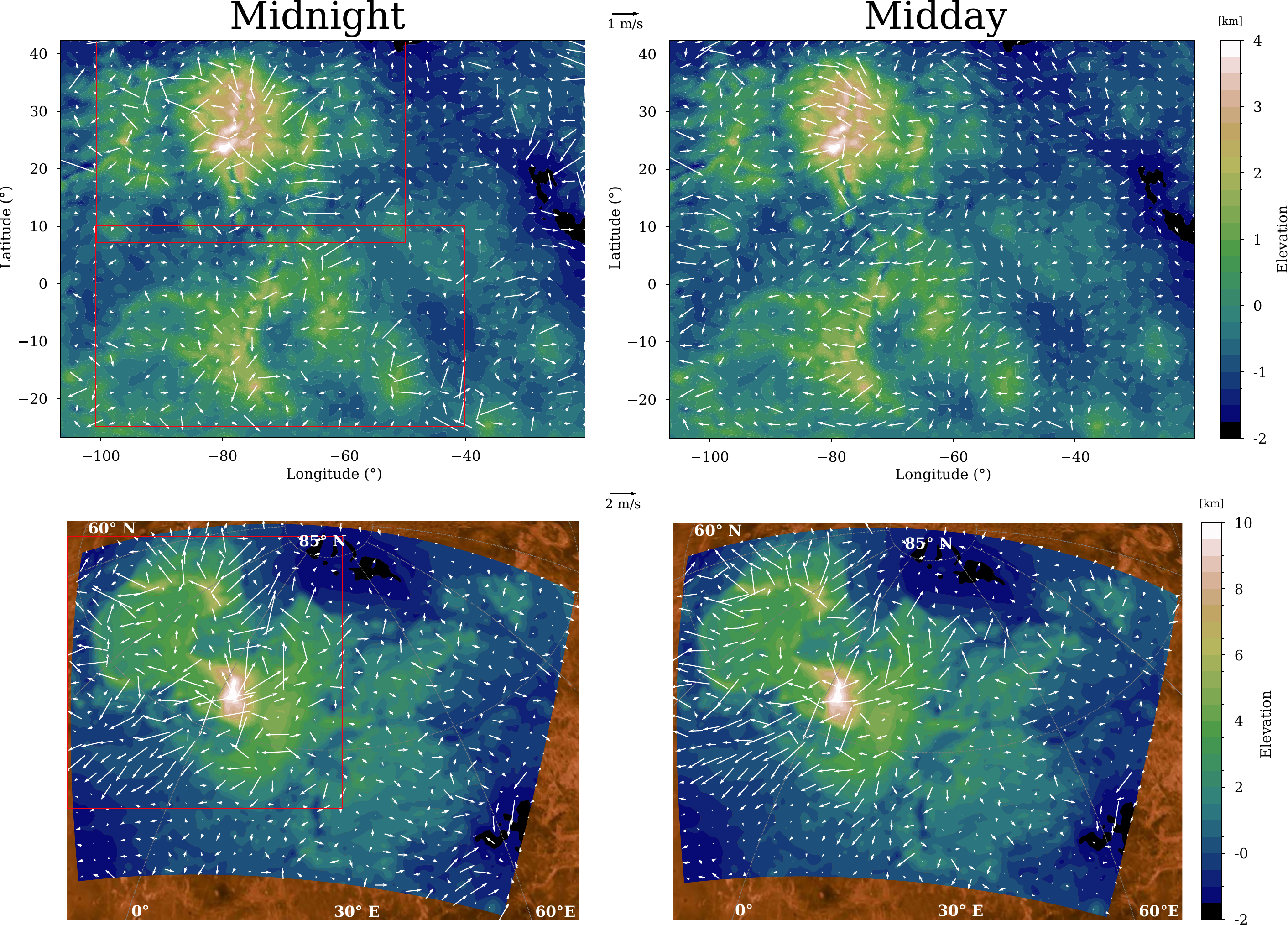}
    \caption{Snapshots maps of horizontal winds 10~m above the local surface (m~s$^{-1}$) at midday (left) and midnight (right) in the centre of the Equatorial (top) and Polar (bottom) domain. Squares are the domains for Phoebe Regio, Beta Regio and Ishta Terra shown in Fig 3.}
  \label{A31}
\end{figure}

\begin{figure}[!ht]
  \centering
  \includegraphics[width=15cm]{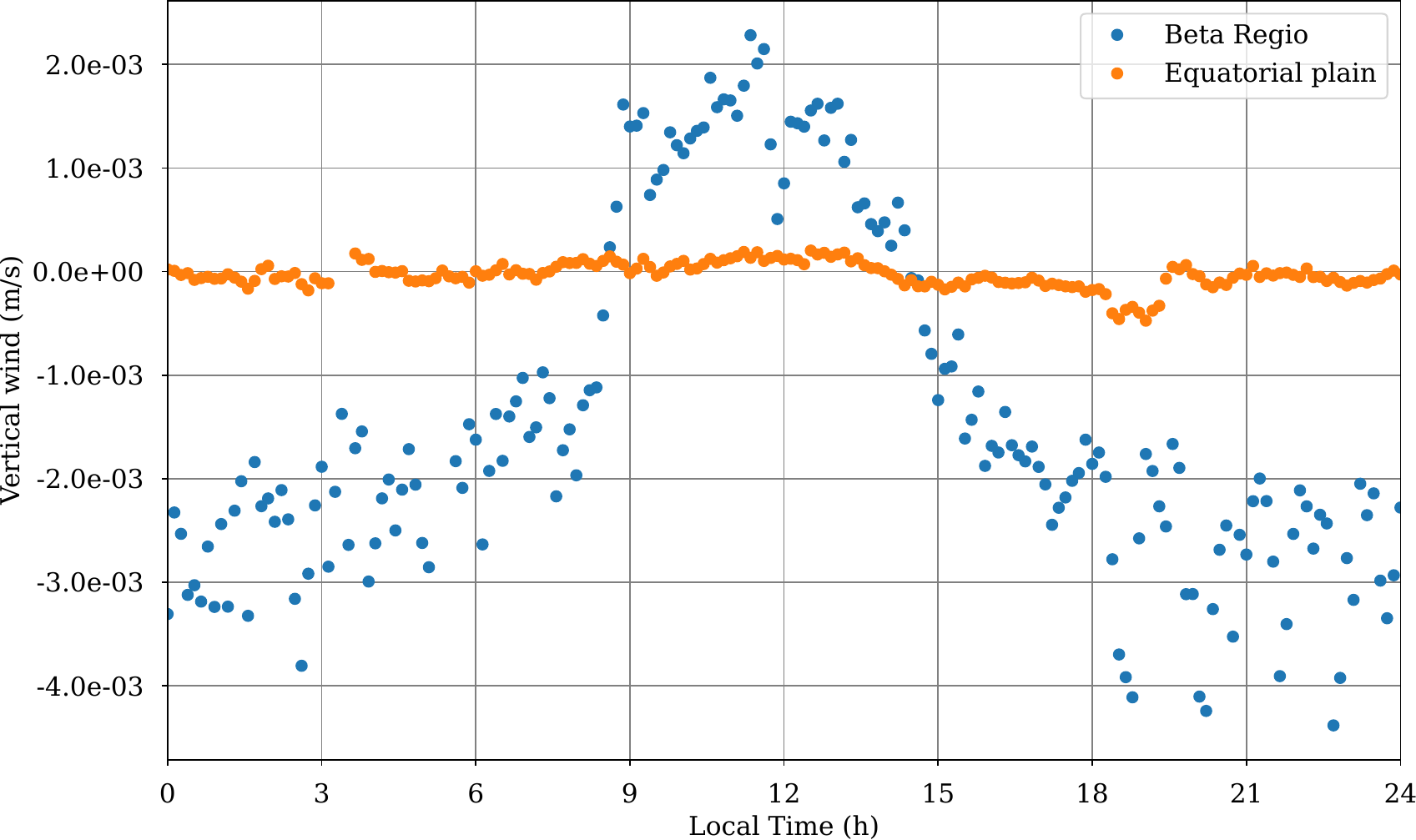}
    \caption{Timeseries of the 2-m vertical wind (m/s) overt Beta Regio and equatorial plain, respectively point I and H on Figure 1.}
  \label{A34}
\end{figure}

\end{appendices}

\end{document}